\documentclass[a4paper,11pt]{article}
\newcommand{\apj}{ApJ}
\newcommand{\aaj}{A\&A}
\newcommand{\apjl}{ApJL}
\newcommand{\mnras}{MNRAS}
\newcommand{\nat}{Nature}
\newcommand{\sovast}{Soviet Ast.}
\newcommand{\prd}{Phys. Rev. D}

\pdfoutput=1 % if your are submitting a pdflatex (i.e. if you have
\usepackage{jcappub} % for details on the use of the package, please
\usepackage[T1]{fontenc} % if needed

\title{\boldmath Multi-epoch study of the gamma-ray emission within the M87 magnetosphere model}

\author{S. Vincent}
\affiliation{DESY,\\Platanenallee 6, 15738 Zeuthen, Germany}
\emailAdd{stephane.vincent@desy.de}
\abstract{
M87 is a nearby radio galaxy that has been detected %is detected 
at energies ranging from radio 
to very high energy (VHE) gamma-rays. 
Its proximity and its jet, misaligned from the line of sight allow detailed 
morphological studies. 
The imaging atmospheric Cherenkov technique (from 100 GeV to 10 TeV) provides 
insufficient angular resolution (few arc-minutes) to resolve the M87 emission 
region. 
However, the short time scale variability observed by MAGIC, HESS and VERITAS 
suggests the TeV emission is coming from a very small region, most likely close to the core. 
We propose that the variable TeV emission may be produced in a pair-starved 
region of the central black hole (BH) magnetosphere, i.e. a region where 
the density of the electron-positron plasma is not sufficient to completely 
screen the accelerating electric field. 
The funnel, a low density and magnetically dominated region around the poles,
appears as a favourable site of low-density where a Blandford-Znajek process 
may explain the main properties of the TeV $\gamma$-ray emission from M87. 
We produce a broadband spectral energy distribution (SED) of the resulting 
radiation and compare the model with the observed fluxes from the nucleus 
of M87, for both low and high $\gamma$-ray activities. 
We finish with a brief discussion on the connection between the accretion 
rate and the intermittence in the formation of gaps in the magnetosphere.
}

\begin{document}
\maketitle
\flushbottom

\section{Introduction}
\label{sec:Introduction}

The Fanaroff-Riley (FR) class I giant radio galaxy M87, situated nearly at the 
center of the Virgo cluster, harbors the first observed extragalactic jet. 
It has been intensively studied at all wavelengths. 
The giant elliptical galaxy is also known to host one of the most massive 
black holes $M_{\mathrm{BH}}\sim6\times 10^9~\mathrm{M}_{\odot}$~\cite{geb09}, 
and a prominent, non aligned jet detectable from radio to X-ray wavelengths. 
Its proximity, $~16.3$ Mpc~\cite{mei07}, makes it one of the two best objects, 
the other being Centaurus A, for testing and understanding extragalactic jets 
of radio-loud Active Galactic Nuclei (AGN) and their powerful engines. 

VHE ($>200$ GeV) gamma-ray emission was first reported by the HEGRA 
collaboration in 1998/99~\cite{aha03}. 
The emission was confirmed by the HESS collaboration in 
2003-2006~\cite{aha06}. 
Additionally, Aharonian \textit{et al.}~\cite{aha06} also reported variability during a high state 
of gamma-ray activity in 2005 with a %on 
timescale of days. 
Timescale for these fluctuations may be related to the characteristic 
time of the central engine
\begin{equation}
\tau \sim M_{\mathrm{BH}} \sim 1/\Omega^{\mathrm{H}},
\end{equation}
where $\Omega^{\mathrm{H}}$ is the angular velocity of the BH, and is in order 
of hours. 
Therefore, the relativistic ejections, responsible for the observed flare, 
appear to be related to the phenomena on the event horizon.
%The observed variability timescale may be compared to the orbital timescale 
%near the innermost stable circular orbit (ISCO) around the BH, 
%i.e. $\Delta t_{\mathrm{min}}=(GM_{\mathrm{BH}}/R_{\mathrm{ISCO}}^3)^{1/2}
%\simeq 7.35\times 10^{-5}~(M_{\mathrm{BH}}/M_{\odot})~\mathrm{s}$, and 
%$R_{\mathrm{ISCO}}=3R_{\mathrm{Sch}}$. 
On the other hand, Chandra X-ray observations revealed a gradual increase 
in activity from HST-1 in the energy range 0.2-6 keV~\cite{har06}. 
The maximum coincided with the TeV flare suggesting HST-1 knot as the more 
likely source of emission. 
However, X-ray timescale analysis presented by Harris \textit{et al.}~\cite{har09} ($\leq 20$ days) 
is still significantly longer than the TeV variability (1-2 days) and 
VLBA observations of HST-1 could not identify substructures down to 
0.15pc~\cite{che07}. 
This upper limit should be compared to the constraint on the emission region 
size from the causality argument, which implies a compact region 
$R_{\mathrm{var}}\leq2\times 10^{-3}~\delta$ pc where $\delta$ is the 
Doppler factor of the relativistically moving source. 
%We can neither claim nor reject the presence of such small features in 
%HST-1 and the core region remains a possible origins of the VHE emission. 
We can neither claim nor reject the presence of such small features in 
HST-1 and the region remains a possible origin of the VHE emission.  
For the nucleus of M87, the data were severely affected by a "pileup" 
effect and are not reliable~\cite{har11} . 
So the open question for the 2005 April TeV flaring episode is could it be that the 
strong flux of HST-1 have suppressed the detection of an enhanced flux coming from 
the X-ray core?

Between January and May 2008, the HESS, MAGIC and VERITAS collaborations 
took part in a joint multi-wavelength monitoring along with VLA and 
Chandra~\cite{bei08}. 
The VHE and Chandra X-ray light curves along with a coincident VLBA radio 
light curve are presented in Acciari \textit{et al.}~\cite{acc09}. 
The flux reached the highest level ever observed, more than 10\% of the 
Crab nebula. 
The observed day scale variability in the energy range between 0.1 TeV 
and 10 TeV, and the fact that the TeV emission appears to be correlated 
with the VLBA jet but not with emission from larger scale and, in particular
HST-1, motivate the consideration that the observed TeV photons originate 
from the BH magnetosphere.

The magnetosphere is a region where a magnetized plasma 
(e.g., an accretion disk and its corona) surrounds a BH and where a wind/jet 
would be generated from the surrounding plasma. 
The Blandford-Znajek mechanism~\cite{bla77} (hereafter BZ), 
used here as a mean of extracting energy from the rotating BH, 
appears to be a plausible mean of harvesting spin energy. 
%exploitation of spin energy. 
Despite some hints~\cite{wil01, mar03} and the general consistency of the 
idea, there is no observational evidence for BH energy extraction. 
In this paper, we study the implication of the activation of the BZ 
mechanism in the magnetosphere and we present $\gamma$-ray spectral 
energy distribution (SED) for quiescent and flaring activities of M87. 
We show that the TeV activity may change significantly depending 
on the injection of the seed charges into the magnetosphere that, in turn, 
may be associated with the accretion rate.

The paper is organized as follows.
In \S 2 existing models for VHE $\gamma$-ray emission from the M87 core are briefly 
reviewed and discussed. 
In \S3 we briefly describe the numerical scheme adopted to simulate the above scenario. 
%%%as well as the Monte Carlo code used here have been described in \citep{vin10}.
In \S 4 we describe the basic model pair production and TeV emission. 
In \$ 5 we discuss the influence of the input parameters.
In \S 6 we show results of Monte Carlo simulations for varying acceleration 
fields. 
In \S 7 we discuss the role of the hole spin parameter on the TeV luminosity.
We summarise and discuss implications for M87 in \S8.

\section{Theoretical scenarios}

Several theoretical scenarios have been invoked to account for the high energies ($>100~\mathrm{MeV}$) 
and VHE $\gamma$-ray emission from M87. 

Georganopoulos \textit{et al.}~\cite{geo05} showed that homogenous one-zone leptonic synchrotron self-Compton (SSC) models are unlikely to explain 
the observed TeV spectrum of M87. 
For an SED whose synchrotron and SSC components peak in the IR and the TeV band, high values of the Doppler factor, $\delta>100$, 
should be inferred. 
A solution to this problem is to assume that the jet emission region is structured, as in the decelerating jet model of refs.~\cite{geo03} and~\cite{geo05}
or as in the spine-layer model \cite{tav08}. 
Both models are characterized by compact sizes of the VHE $\gamma$-ray emission %that may account for the 
%the short-timescale variability
and no obvious correlation between the low and high-energy emission 
(for instance the keV component will not exactly follow the TeV component). 
%%A feature of those models is that 
They also predict very strong gamma-ray absorption and relatively soft spectrum, and hard spectra are difficult to achieve.

Lenain \textit{et al.}~\cite{len08} present a modification of the one-zone SSC model. 
In that work, the bulk of the observed emission is proposed to be produced in compact sub-volumes of the main outflow. 
Such a scenario could in principle account for VHE high states; 
the individual blob radii of about $10^{14}~\mathrm{cm}$ is compatible with the short variability and 
the blobs are ejected from the core with reasonable bulk velocities.
However, a low magnetic field in the emitting region is required which may be in disagreement with the fact that these regions of the jet are 
likely strongly magnetized. 
%%%In the work by \cite{gia10},
Giannios \textit{et al.}~\cite{gia10} consider mini-jets of energetic particles moving relativistically within the jet. 
Depending on the direction of motion of the mini-jet, different variability patterns and various variability timescales could, in principle, 
be expected.  

Recently, a different type of modelling has been brought forward to explain the observed VHE $\gamma$-ray emission. 
In ref.~\cite{bar10}, the authors proposed a scenario where a red giant star interacts with the base of the jet. 
The VHE emission is due to the interaction of the protons from the jet with the disrupted red giant envelope. 
To accommodate the light curve and the energy spectrum of the April 2010 flare, a jet power of  
$\sim10^{44}~\mathrm{erg.s}^{-1}$ collimated into a narrow jet with semi-opening angle $\theta\sim2^{\circ}$ 
are assumed. 
At the distance from the hole at which the cloud crosses the jet $\sim10^{16}~\mathrm{cm}$, such a narrow jet 
would not concur with the VLBI observations showing that the jet opening angle reach $\sim 60^{\circ}$ within 
1 milli-arcsecond of the core~\cite{jun99, ly07, asa12}.

The compactness of the particle accelerators operating in the vicinity of the BH and the absence of a significant cutoff 
in the spectrum imply that the particle acceleration mechanism is highly efficient. 
Magnetospheric models \citep{ner07, rie08, vin10, lev11}, where the non-thermal particle acceleration and emission processes occur 
at the base of the magnetosphere may then become particularly interesting. 

\section{Outline of the simulation}

The reader should refer to \cite{vin10} for the details of the simulation.  
We only give a brief overview. 
We carried out a Monte Carlo simulation of an electromagnetic cascade in a BH magnetosphere. 
The simulations use the exact cross-section of the full electron-photon cascade involving photon-photon pair production interactions 
and inverse Compton scattering. 
The number density of the target photon field is uniformly distributed into an infrared emission region of radius $R_{\mathrm{IR}}$ around 
the BH and the energy distribution is approximated as a $\delta$ function at energy $E_{\mathrm{IR}}$. 
The electrons are assumed to be initially at rest with respect to the locally inertial observer and injected at the base of the gap. 
%%%For the charged particle and 
First, the mean interaction length is calculated, and then an interaction length is sampled from an exponential distribution with this mean. 
At the interaction point, the interactions are treated as described in Sections 3.3 and 3.5 in \cite{vin10}. 
At each step of integration of the electron's trajectory, the energy and power of synchrotron emission are calculated. 
If the interaction length is greater than $R_{\mathrm{IR}}$ or if the condition on the pair energy threshold for a head-on collision is not satisfied, 
the energy of the particle is printed out. 

\section{Properties of the central AGN}

In what follows, we review the non-thermal particle acceleration 
and VHE emission processes that may occur in the vicinity of 
magnetized supermassive BH. 
We discuss the internal $\gamma$-ray opacity and the consequences 
for the TeV observations of M87. 
And, we finally note that the intermittence of the gap formation 
may be related to the changes in the accretion rate.

\subsection{Particle acceleration close to the central black hole}
In this scenario, the Kerr BH is threaded by an accretion-deposited 
magnetic field. 
The charged particles are electromagnetically accelerated to extreme 
relativistic velocities in a region with a potential drop created 
by the inefficiency of charge supply (BZ, \cite{mac82}).
The accelerated particles are usually postulated to be leptons~\cite{bes92, hir98} 
but protons have also been proposed~\cite{lev00, aha02}.
The spectral and timing characteristics of the VHE emission are 
associated to the physics of the processes taking place close to the
supermassive BH. 
The mass of the BH in M87 is well established and the expected minimal
timescale variability for a non-rotating BH is 
$\Delta t=2GM_{\mathrm{BH}}/c^3\simeq 6\times 10^4~\mathrm{s}\simeq 1~\mathrm{day}$, 
and the observed short timescale variability of TeV emission of days
indicates that the emission originates within the last stable orbit
unless the radiation is produced in a relativistic bulk motion.

In a stationary plasma-filled magnetosphere, the structure of the
electromagnetic field satisfies the condition $\mathbf{E}.\mathbf{B}=0$,
where $\mathbf{E}$ and $\mathbf{B}$ are the electric and magnetic field
respectively. 
Around a rotating BH, the electric field arises as a consequence of
the coupling of the magnetic field to the frame dragging effect, see
e.g.~\cite{wal74}. 
The characteristic charge density $\rho_{\mathrm{GJ}}$ needed to
neutralize the parallel component of the electric field in the
magnetosphere is the so-called Goldreich-Julian density~\cite{gol69}. 
If the local value of the charge density $\rho_e$ differs
significantly from $\rho_{\mathrm{GJ}}$ in any region, a gap forms and
an electric field with a component parallel to the magnetic field
provides conditions for particle accelerations to arise. 
%The energy change of particles of charge $e$ accelerated along $\Delta l$ in
The change in energy of charged particles accelerated along $\Delta l$ in
a region of the size of the gravitational radius
$R_{\mathrm{G}}=GM_{\mathrm{BH}}/c^2$ is
\begin{equation}
\Delta E\simeq 4.4\times 10^{20} \kappa
\left(\frac{B}{10^4\mathrm{G}}\right)
\left(\frac{M_{\mathrm{BH}}}{10^9\mathrm{M}_{\odot}}\right)
\left(\frac{\Delta l}{h}\right)~\mathrm{eV},
\end{equation}
%$h$ is the gap height of a size of order $R_{\mathrm{G}}$ and $\kappa$
%is the efficiency of the acceleration mechanism, it is related to the
%deviation from the ideal case, i.e. when $\mathbf{E}.\mathbf{B}$ differs
%from zero but stays close to zero and how pairs can screen out the
%electric field.
where $h$ is the gap height of a size of order $R_{\mathrm{G}}$.
The efficiency of the acceleration mechanism $\kappa$ is related to the
deviation from the ideal case, i.e. when $\mathbf{E}.\mathbf{B}$ can differ
from zero but stays close to zero and how pairs can screen out the
electric field.

Radiative energy losses always compete with acceleration and introduce
additional constraints on the maximum achievable particle energies. 
In the hypothesis of the electronic origin of the TeV $\gamma$-ray
emission in terms of high-energy electrons, the main emission
mechanisms are synchrotron radiation and inverse Compton (IC)
scattering. 
General relativistic magnetohydrodynamics simulations show that large
scale radial magnetic field may form near the BH, e.g.~\cite{haw06, bec09}, 
and the energy losses of electrons are not expected to be
dominant over curvature radiation.
In the potential gap, the motion of the electron-positron pairs along
the magnetic field is maintained by the $E_{||}$ field. 
The momentum component perpendicular to the magnetic field $\mathbf{B}$,
owing to the IC scattering with the ambient photons, is rapidly lost,
and the pitch angle is almost null. 
Therefore, the synchrotron radiation originates from the gyration of
the charged particles in the component of the magnetic field
$\mathbf{B}$. 
%The requirement of balance between the synchrotron energy loss and the
%acceleration rate gives the maximum Lorentz gamma factors
The requirement that the synchrotron energy loss is balanced with the
acceleration rate gives the maximum Lorentz gamma factors
\begin{equation}
\gamma^{\mathrm{max}}_{\mathrm{sync}}\simeq 1\times 10^8\left
(\frac{B}{1\mathrm{G}}\right)^{-1/2}\kappa^{1/2}.
\end{equation}
For IC scattering, the maximum energy of $e^{\pm}$ pairs is
\begin{equation}
\gamma^{\mathrm{max}}_{\mathrm{IC}}\simeq 2\times 10^7\left
(\frac{B}{1\mathrm{G}}\right)^{1/2}\left(\frac{R_{\mathrm{IR}}}{R_{G}}\right)\kappa^{1/2},
\label{ICestimate}
\end{equation}
where $U_{\mathrm{rad}}=L_{\mathrm{IR}}/(4\pi R^2_{\mathrm{IR}}c)$ is
the energy density of radiation, $L_{\mathrm{IR}}$ is the luminosity
of the source coming from matter accreting onto the hole and for the
infrared luminosity $L_{\mathrm{IR}}=10^{41}$ erg.s$^{-1}$. 
Eq. (\ref{ICestimate}) is determined under the assumption that the IC scattering takes place 
in the Thomson regime. 
The highest-energy electrons upscatter the background radiation into the Klein-Nishina regime, 
in which the efficiency of the IC scattering is reduced. 
A proper accounting of the decrease in IC loss efficiency would result in higher electron energies. 

For an explanation of the observed TeV $\gamma$-ray radiation by IC scattering, one should assume 
that the energy losses of electrons are not dominated by synchrotron radiation. 
We can set an upper bound on the value of the magnetic field strength in the infrared region 
by equating the energy densities of the magnetic field and of radiation, 
i.e. $U_{\mathrm{mag}}=U_{\mathrm{rad}}$, and
\begin{equation}
B=\sqrt{\frac{L_{\mathrm{IR}}}{2cR_{\mathrm{IR}}}}\sim 0.1\mathrm{G}.
\end{equation}
The size of the infrared emission region is taken to be $R_{\mathrm{IR}}=15R_{\mathrm{G}}$, 
and if $B< 0.1\mathrm{G}$, the absorption of the first-generation of $\gamma$-rays will trigger an 
electromagnetic cascade.

\subsection{Escape of the $\gamma$-ray photons}
In this section, we examine the conditions under which the
$\gamma$-ray photons in the energy range covered by VERITAS
($E=0.1-10\mathrm{TeV}$) escape from the regions where they are
produced. 
The spectrum of the VHE photons observed by air Cerenkov detectors
and the assumption that these photons originate from the
magnetosphere imply that the pair production opacity at these energies
does not exceed unity. 
The highest energy $\gamma$-rays produce pairs in interaction with the
soft photon background inside the compact source. 
The cross-section of photon-photon pair-production depends on the
center of mass energy of the colliding photons,
\begin{equation}
s=\frac{E\epsilon(1-\cos\theta)}{2\mathrm{m}_e^2c^4},
\end{equation}
where $E$ and $\epsilon$ are the energies of the photons and $\theta$
is the colliding angle. 
The dominant source of IR-optical photons that can pair create with
the TeV photons is the ambient nucleus emission. 
The observed luminosity of the M87 nucleus at $\nu=10^{13}-10^{14}$ Hz
is $\mathrm{L}_{\nu}\sim 10^{41}~\mathrm{erg.s}^{-1}$~\cite{kha04, why04}. 
Starting from the threshold condition at $s=1$, the pair-creation
cross section peaks at $\sigma_{\gamma\gamma}\sim
0.2\sigma_{\mathrm{T}}\sim 1.33\times 10^{-25}~\mathrm{cm}^2$ for
photons with total energy $s\sim3.5-4$ in the center-of-mass frame, 
and then decreases as $s^{-1}\log s$. 
TeV photons interact mostly with target soft photons of energy
$\epsilon\sim 1~\mathrm{eV}$. 
The number density of soft photons is
$n_{\mathrm{IR}}=\mathrm{L}_{\mathrm{IR}}/4\pi
R_{\mathrm{IR}}^2c\epsilon_{\mathrm{IR}}$, where
$\mathrm{L}_{\mathrm{IR}}\sim 10^{41}\mathrm{erg.s}^{-1}$. 
Thus the annihilation optical depth is
\begin{equation}
\tau_{\gamma\gamma}\simeq 20
\left(\frac{\mathrm{L}_{\mathrm{IR}}}{10^{41}\mathrm{erg.s}^{-1}}\right)
\left(\frac{R_{\mathrm{IR}}}{R_G}\right)^{-1}
\left(\frac{E}{1~\mathrm{TeV}}\right).
%\left(\frac{\epsilon_{\mathrm{IR}}}{1~\mathrm{eV}}\right)^{-1}.
\end{equation}
The considered region would be transparent to the $\sim 1\mathrm{TeV}$
photon if $R_{\mathrm{IR}}>20 R_G$. 
Although infrared telescopes prevent us from resolving the size of the
radiation source, HST spectroscopic observations constraints the size
of the nuclear disk in M87 to $\sim 6~\mathrm{pc}\sim 2\times 10^4 R_G$~\cite{mac97}. 
M87's proximity makes the absorption of $\gamma$-rays in the cosmic
background radiation field insignificant.

\subsection{The role of the inefficient accretion flow}
\label{section:The role of the inefficient accretion flow}
As explained above, two properties of the central engine have some
profound implications on the possible mechanisms of VHE $\gamma$-ray
emission: the plasma density and the magnetic field environment. 
Matter density near the event horizon may be estimated from the hot
interstellar medium at the Bondi radius, i.e. the radius at which the
gravitational potential of the BH begins to dominate the dynamics of
the hot gas and the gas starts to fall into the BH. 
Henceforth, we use the conventional notation that expresses the
accretion rate in units of the Eddington rate
$\dot{m}=\dot{M}/\dot{M}_{\mathrm{Edd}}$, where $m$ is a scale factor
and the Eddington accretion rate
$\dot{M}_{\mathrm{Edd}}L_{\mathrm{Edd}/}\eta c^2$, is the rate
necessary to sustain the Eddington luminosity. 
The accretion efficiency factor is noted $\eta$ and $\eta\sim 10\%$. 
Di Matteo \textit{et al.}~\cite{dim03} gives an estimate of the Bondi accretion rate of the
nucleus of M87 using the Chandra observations.  
The corresponding value is an upper limit of the accretion rate,
$\dot{M}<\dot{M}_{\mathrm{Bondi}}\simeq
0.1~\mathrm{M}_{\odot}.\mathrm{yr}^{-1}$. 
The Bondi accretion rate in this units is
$\dot{m}_{\mathrm{Bondi}}=1.6\times 10^{-3}$. 
Interestingly, this indicates that the accretion proceeds in a
radiatively inefficient way and suggest that the density of the
injected charges may not exceed the Goldreich-Julian charge density
$\rho_{\mathrm{GJ}}$ in at least some region of the magnetosphere. 
Consequently, a complete screening of the aligned electric field
cannot be accomplished and a gap forms. 
Whereas at sufficiently large accretion rates, the seed charges
density may exceed $\rho_{\mathrm{GJ}}$, giving rise to complete
screening and leads to the formation of a force-free outflow
($\mathbf{E}.\mathbf{B}=0$). 

Independently from the uncertainty on the accretion regime, one can
estimate the ion density near the BH,
\begin{equation}
\label{n_ion}
n_{\mathrm{ion}}=\frac{\dot{M}}{2\pi r^2 m_p v_r}\simeq 10^7~\mathrm{cm}^{-3},
\end{equation}
where $v_r$ is the radial velocity. 
In ref.~\cite{nar95}, the authors show that at lower radii ($r<10^2\times
R_{\mathrm{Sch}}$), the electron temperature becomes large, $T_e\sim
10^9~\mathrm{K}$, and we assume in equation (\ref{n_ion}) that the
accretion material moves at relativistic speeds, i.e. $v_{r}\simeq c$.

The magnetic field strength in the vicinity of the BH can be related
to the accretion rate $\dot{m}$. 
We assume that the accreting gas is roughly in equipartition with the
magnetic field, and we write the total pressure $p$ as
\begin{equation}
p=p_g+p_m,\quad p_g=\beta p,\quad p_m=1-\beta p,
\end{equation}
where $p_g$ is the gas pressure, $p_m$ the magnetic pressure, and we
take $\beta=1/2$. 
The equipartition magnetic field is then determined from the magnetic
pressure, i.e, $B^2_{\mathrm{eq}}/8\pi=0.5\rho_e c^2$, and yields
\begin{equation}
B_{\mathrm{eq}}\simeq
370~\dot{m}^{1/2}\left(\frac{R_{\mathrm{Sch}}}{r}\right)\sim 15~\mathrm{G},
\end{equation}
giving a maximum acceleration energy
\begin{equation}
E_{\mathrm{max}}=eB_{\mathrm{eq}}R_{\mathrm{Sch}}\sim 10^{18}~\mathrm{eV}.
\end{equation}
The Goldreich-Julian density\footnote{The given expression is the
  flat-space Goldreich-Julian charge number. A better estimate for
  $n_{\mathrm{GJ}}$ would use the simulation derived currents density
  $J^{\mu}$ and the relation $n_{\mathrm{GJ}}\equiv J^{\mu}U_{\mu}/e$, 
where $U_{\mu}$ is the 4-velocity according to an observer at
infinity.} $n_{\mathrm{GJ}}$ can be related to the accretion rate
\begin{equation}
\label{n_GJ}
n_{\mathrm{GJ}}\simeq \frac{\Omega B_{\mathrm{eq}}}{4\pi ec}=4\times
10^{-4} \dot{m}^{1/2}~\mathrm{cm^{-3}}.
\end{equation}
Mo{\'s}cibrodzka \textit{et al.}~\cite{mos11} gave an estimate of the charge density
for a monopole magnetosphere taking into account general relativistic
effects. 
We have checked that our flat spacetime estimate is consistent with
the Kerr spacetime estimate. 
For a maximally rotating BH with mass $10^9~\mathrm{M}_{\odot}$, the
angular rotation speed is $\Omega=d\Phi/dt=1/2 M_{\mathrm{BH}}\sim
9\times 10^{-5}~\mathrm{rad.s}^{-1}$. 
The value of $n_{\mathrm{GJ}}$ derived from equation (\ref{n_GJ}) makes
the possibility of creating gaps in the accretion disk or on the
plunging region\footnote{The plunging region lies between the
  innermost stable circular orbit and the event horizon} unlikely. 
However, magnetohydrodynamics simulations of models of zero obliquity
BH accretion, in which the accretion flow angular momentum is parallel
to the BH spin, exhibit a low density evacuated funnel region near the pole. 
This region looks favourable for the formation of gaps. 

\begin{figure}[!h]
\centering
\includegraphics[width=0.3\textwidth]{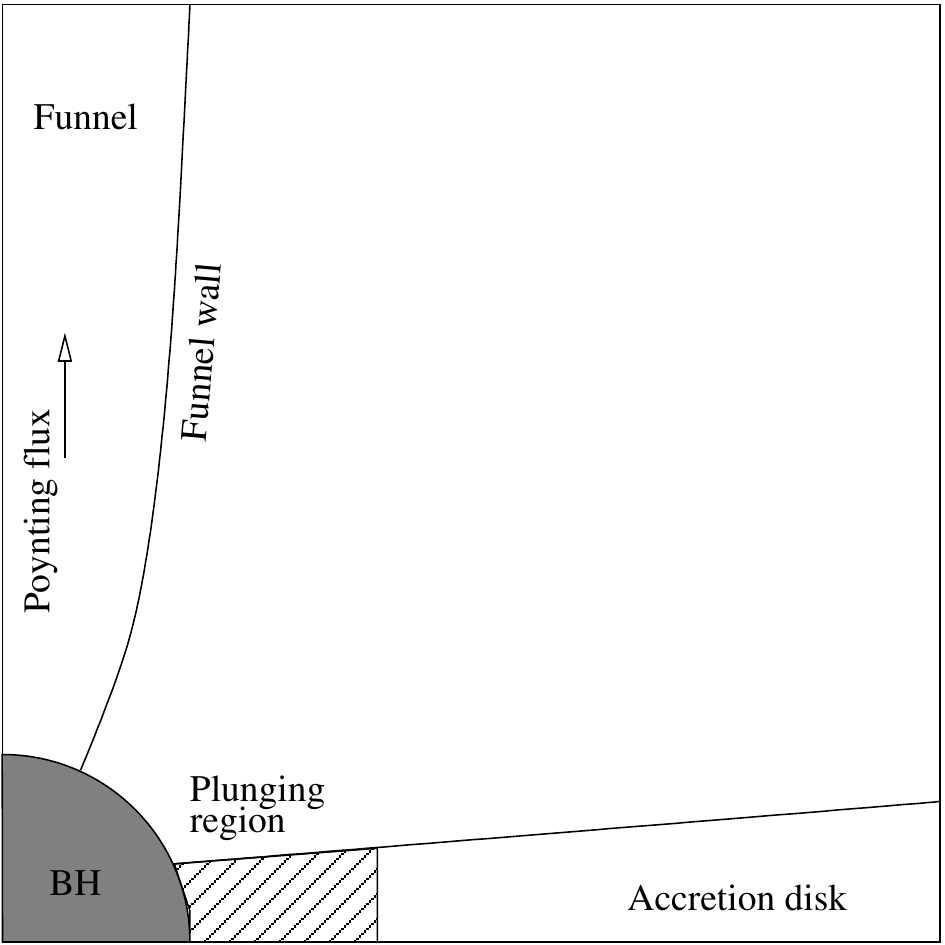}
\caption{\label{m87} Schematic illustration of the main dynamical features and locations of the black hole magnetosphere. 
They are the accretion disk, a matter dominated region; the plunging region; 
the funnel, a magnetically dominated region, where the magnetic field transports energy outward in the form of a Poynting flux.}
\end{figure}

In ref.~\cite{mck04}, the authors showed that the funnel region is nearly force-free and is well 
described by the stationary force-free magnetosphere model. 
But an important question remains: how to populate the funnel? 
Direct feeding of the funnel with plasma seems unlikely as charged particles 
would have to cross magnetic field lines. 
Moreover, funnel magnetic field lines are almost radial near the event horizon 
and then run in a smooth helicoidal shape, and do not intersect the disk~\cite{kom04, haw06}. 
These structures are illustrated in figure \ref{m87}.
 
Levinson \& Rieger~\cite{lev11} propose a mechanism to feed the polar region by annihilation of MeV 
photons\footnote{such a scenario favours jet made pairs over jets made of electron-ion plasma, 
and restricts the production of neutrinos in AGN jets.} that are emitted by the hot gas near the horizon. 
The corresponding number of pairs is 
\begin{equation}
\label{n_pm}
n_{\pm}\simeq 5.5\times 10^7~\dot{m}^4~\mathrm{cm}^{-3}
\end{equation}
and
\begin{equation}
n_{\pm}/n_{\mathrm{GJ}}\simeq 10^{11}~\dot{m}^{3.5},
\end{equation}
below an accretion rate $\dot{m}\leq 7\times 10^{-4}$, the injection of charges cannot provide complete 
screening of the magnetosphere. 
The dependence on the injection rate suggests that a gap may form during periods of low accretion 
and the emission from the gap may be intermittent.

%%%\section{Structure of the potential gap}
\section{Influence of the model parameters}
In this section, we study the influence of several model parameters, such as the magnitude of the induced electric field. 
In the analysis of our results, we focus on the expected gamma-ray spectral signatures in a region of parameter space particularly well suited to reproduce the broadband spectral energy distributions of the radio galaxy M87. 
In each one of our simulations, we have assumed an infrared luminosity of $\mathrm{L}_{\nu}\sim 10^{41}~\mathrm{erg.s}^{-1}$
at frequencies $\nu=10^{13}$ Hz. 
The parameters used for the individual runs are quoted in Table~\ref{tab:parameters}.

%\subsection{Steady axisymmetric equations of electrodynamics in absolute space.}
%\subsection{Basic equations}
\subsection{Structure of the potential gap}
%In the (3+1) formalism, the hole is characterized by the universal time coordinate $t$, measured 
%by a clock at infinity, and a absolute three-dimensional space with curved geometry. 
For a rotating Kerr BH of mass $M$ and angular $J$, the Boyer-Lindquist spatial coordinate 
system is, with a parameter $a=J/Mc$
\begin{eqnarray}
ds^2&=&(\rho^2/\Delta)dr^2+\rho^2d\theta^2+\varpi^2d\phi^2,\\
\rho^2&=&r^2+a^2\cos^2\theta,~\Delta=r^2-2GMr/c^2+a^2,~\Sigma^2=(r^2+a^2)^2-a^2\Delta\sin^2\theta,\\
\varpi&=&(\Sigma/\rho)\sin\theta,~\omega=2aGMr/c\Sigma^2,~\alpha=\rho\Delta^{1/2}/\Sigma
\end{eqnarray}
where $\omega$ is the angular velocity of the local zero angular momentum observer (ZAMO) with
respect to infinity, representing the dragging of the local inertial frame, $\alpha$ is the lapse function, 
or the redshift factor between the ZAMO and infinity.\\
We consider a stationary axisymmetric flow of cold plasma around a rotating black hole.
Within this approach the magnetic field is expressed by
\begin{equation}
\bf{B}=\frac{\nabla\psi\times\bf{e_{\hat{\phi}}}}{2\pi\varpi}-\frac{2I}{\alpha\varpi},
\end{equation}
with $\psi(r, \theta)$ is the magnetic flux function determining the poloidal magnetic field and 
$I(r, \theta)$ is the electric current producing the toroidal magnetic field. 
The 3-gradient of a scalar function $f(\bf{x})$=$f(r, \theta, \phi)$ in the Boyer-Lindquist coordinates is
\begin{equation}
\nabla f=\frac{\sqrt{\Delta}}{\rho}(\partial_{r}f)\bf{e_{\hat{r}}}+\frac{1}{\rho}(\partial_{\theta}f)\bf{e_{\hat{\theta}}}
+\frac{1}{\varpi}(\partial_{\phi}f)\bf{e_{\hat{\phi}}}.
\end{equation}

While the precise structure of the BH field is unclear, steady-state structure of spinning BH magnetosphere 
has been obtained through time-dependent general relativistic force-free numerical simulations 
(e.g.~\cite{kom02, mck06, con13}). 
These works found that the numerical solution evolves toward a solution which is very close to the monopole solution 
\begin{eqnarray}
\psi=\psi_0(1-\cos\theta),\quad I=\frac{\Omega_{F}\psi_0}{4\pi}\sin{^2}{\theta},
\end{eqnarray}
$\Omega_{F}=\Omega_{H}/2$ is called the angular velocity of magnetic field, and $\Omega_{H}$ is the angular 
velocity of the BH. 

\begin{figure}[tbp]
\centering % \begin{center}/\end{center} takes some additional vertical space
\includegraphics[width=\textwidth]{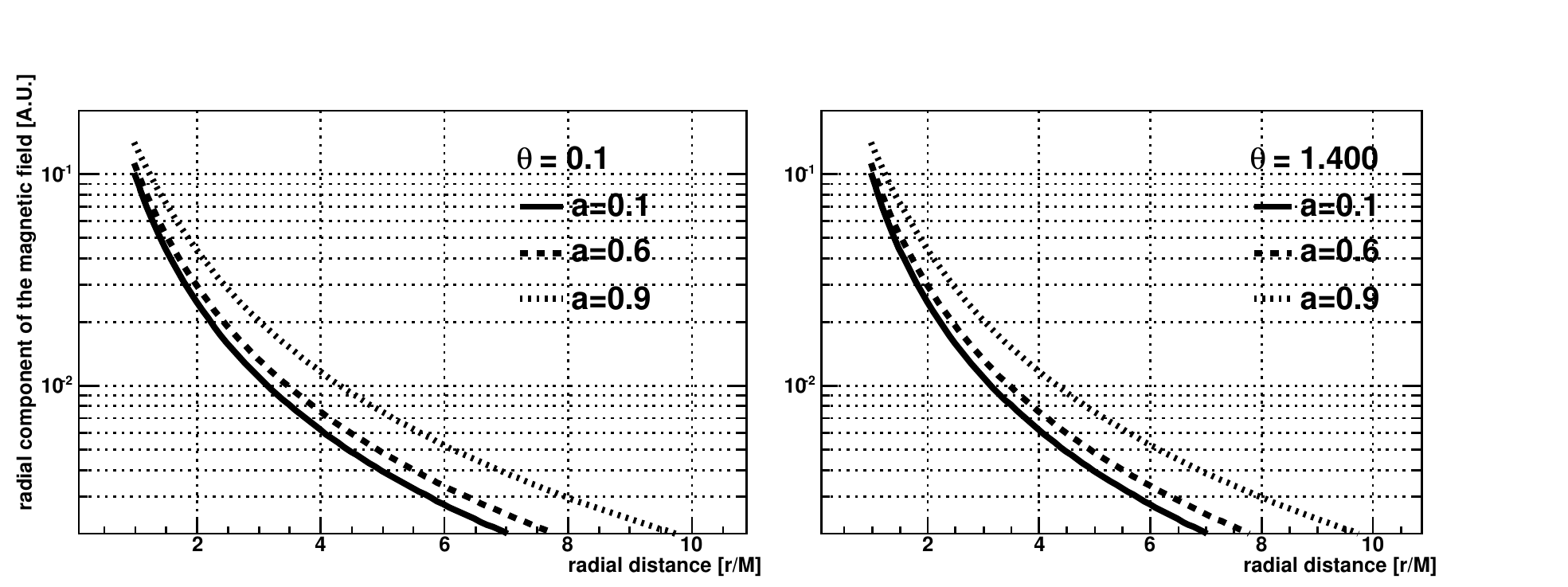}
\hfill
%\vspace{-0.65cm}
%%%\includegraphics[width=\textwidth]{plots/azimuthalComponentBField.eps}
\includegraphics[width=\textwidth]{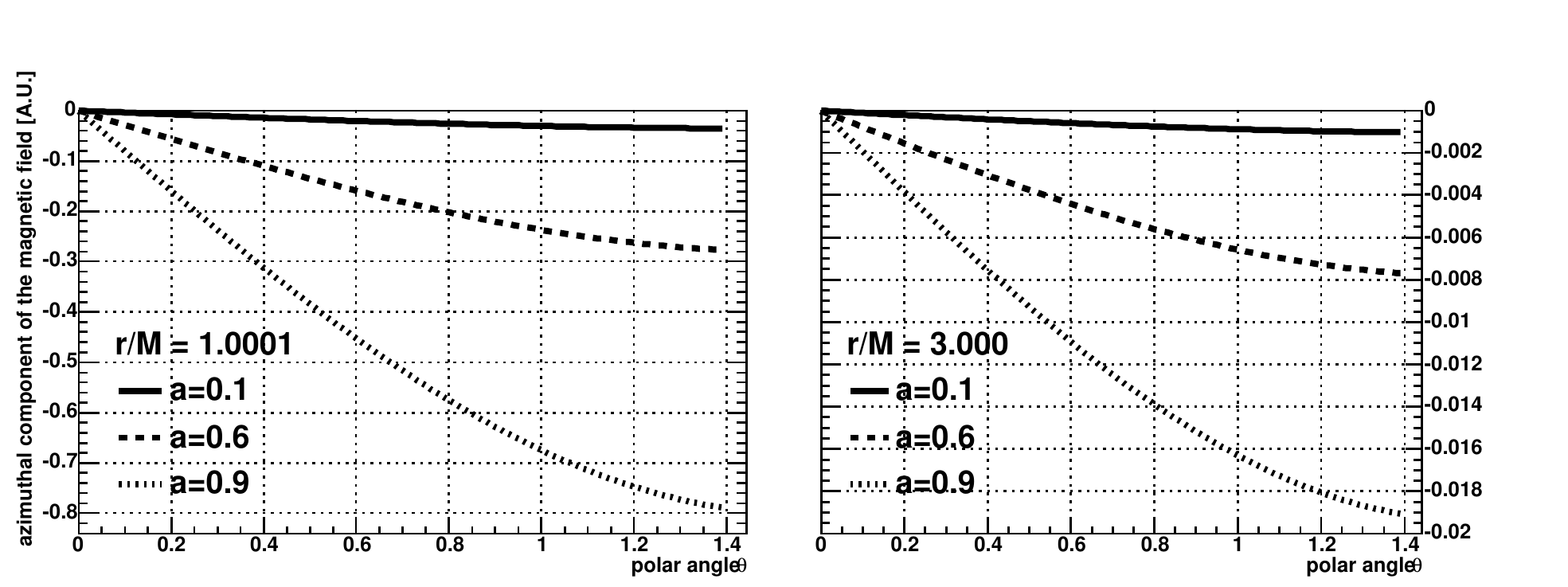}
% "\includegraphics" is very powerful; the graphicx package is already loaded
\caption{\label{fig:bField} Sequence of shapes of the radial and azimuthal components of the magnetic field lines with increasing spin parameter $a$. }
\end{figure}

In Fig.~\ref{fig:bField},  we show the characteristic magnetic field configuration $B_{\phi}$ (top panel) and $B_{r}$ 
(lower panel) for $a=0.9,~0.6,~0.1$ near the BH horizon. 
Near the hole the magnitude of the potential drop along the $\bf{B}$ field lines is $V\sim aB$.

\subsection{Features of the non-thermal emission}
\begin{figure}[tbp]
\centering % \begin{center}/\end{center} takes some additional vertical space
\includegraphics[width=\textwidth]{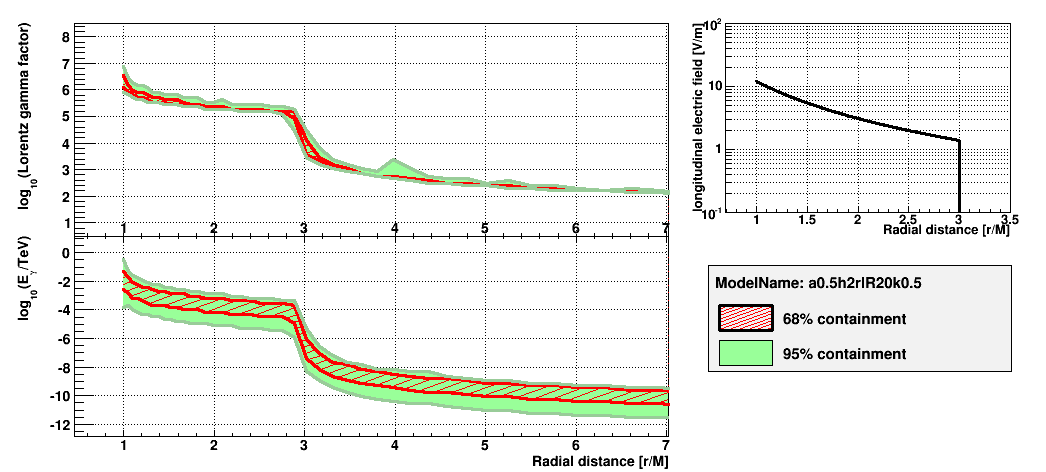}
\caption{\label{fig:LorentzFactor} Radial profiles of the Lorentz factor of the charged particles (\textit{top panel}) and of the photon spectra (\textit{bottom panel}). On every panel, the shaded areas show the 68\% and the 95\% containment range. 
}
\end{figure}

Fig.~\ref{fig:LorentzFactor} illustrates the evolutions of the accelerated particles along the field line, which penetrate the gap.   
The top and the bottom panels show the evolutions of the Lorentz factor of the particles and the energy of the Comptonized photons. 
The shaded bands indicate the 95\% and the 68\% containment. 
The inset in the top panel of Fig.~\ref{fig:LorentzFactor} shows the variation of the mean value of the accelerating electric field as a function of the radial distance. 
A charged particle introduced at the base of the gap is rapidly accelerated by the electric field and is boosted above $10^{6}$ on the Lorentz factor. 
This particle escape from the gap at $r/M_{\mathrm{BH}}=3$ where the accelerating field vanishes, and rapidly lose their energy via the Compton scattering. 
%Because the cooling length of the Compton scattering, $l_{\mathrm{IC}}/h\sim3\times10^{-12}(R_{\mathrm{IR}})^2(h\gamma)^{-1}$ becomes comparable to the gap %height at the Lorentz factor of $\gamma\sim10^{6}$, the particles .

%%%The effect of an increasing particle acceleration size - corresponding to a longer observed variability timescales, $t_{\mathrm{var.}}$, is illustrated in Fig.~\ref{fig:ModelComparison1}. 
%%%An interesting qualitative change of the SED can be seen; while for a vacuum gap of size $h\sim2M_{\mathrm{BH}}$ corresponding to $t_{\mathrm{var.}}\sim$, the   

\begin{figure}[tbp]
\centering % \begin{center}/\end{center} takes some additional vertical space
\includegraphics[width=\textwidth]{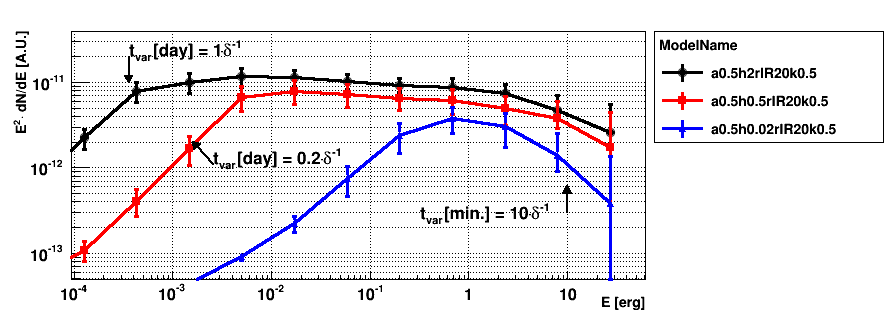}
\includegraphics[width=\textwidth]{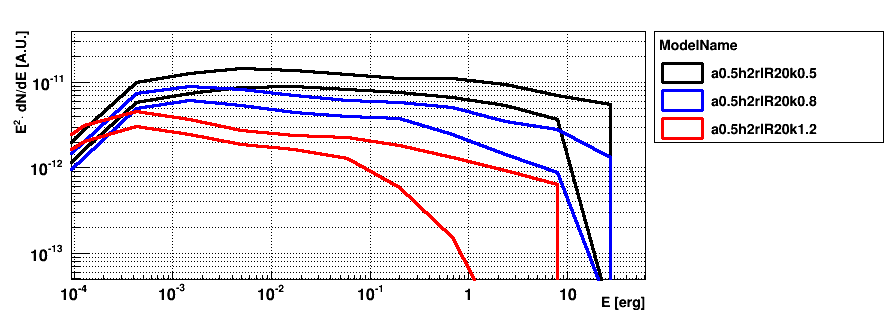}
\caption{\label{fig:ModelComparison1}Top: Photon spectra for different size of the gap $h$. The timescale variability $\mathrm{t}_{\mathrm{var}}$ associated to the size of the gamma-ray emitting region is also indicated. The lines represent the corresponding averaged SED. 
Bottom: Photon spectra for different intensities of the accelerating electric field, parametrised by the parameter $\kappa$. The bands show the spread in the spectral energy distributions (SED). 
}
\end{figure}

An interesting qualitative change of the SED is illustrated in Fig.~\ref{fig:ModelComparison1}. 
The figures indicate that the peak luminosity of the gamma-ray spectra associated to inverse Compton emissions is related to the size of the vacuum gap. 
For example, a gap of size $h\sim2M_{\mathrm{BH}}$ corresponding to a variability on timescale of $t_{\mathrm{var.}}\simeq1$~day, produce a peak at GeV energy.   
For a thinner gap of $h\sim0.02M_{\mathrm{BH}}$ corresponding to a timescale of $t_{\mathrm{var.}}\simeq10$~mins, the peak luminosity is at hundredth of GeV.

Subsequently, we investigate the influence of changing the value of the $\kappa$ parameter that determines the strength of the electric field. 
Fig.~\ref{fig:ModelComparison1} illustrates the hardening of the photon spectra at GeV and TeV energies with an increase of the magnitude of the accelerating field. 
%The synchrotron energy bump at keV to MeV energies peaks significantly at higher energy during    

%%%\begin{figure}[tbp]
%%%\centering % \begin{center}/\end{center} takes some additional vertical space
%%%\includegraphics[width=1.\textwidth]{plots/modelComparison4.png}
%%%\caption{\label{fig:ModelComparison} Photon spectra for different intensities of the accelerating electric field, parametrized by the parameter $\kappa$.The shaded bands show the spread in the spectral energy distributions (SED). 
%%%The dotted lines represent the corresponding averaged SED.}
%%%\end{figure}

\begin{table}[tbp]
\centering
\begin{tabular}{|lcccc|}
\hline
ModelName & a/M$_{\mathrm{BH}}$ & h/R$_{\mathrm{G}}$ & R$_{\mathrm{IR}}/$R$_{\mathrm{G}}$ & $\kappa$ \\
\hline
a0.5h2rIR20k0.5 & 0.5 & 2.0 & 20.0 & 10$^{-0.5}$ \\
a0.5h2rIR20k0.8 & 0.5 & 2.0 & 20.0 & 10$^{-0.8}$ \\
a0.5h2rIR20k1.2 & 0.5 & 2.0 & 20.0 & 10$^{-1.2}$ \\
a0.5h0.5rIR20k0.5 & 0.5 & 0.5 & 20.0 & 10$^{-0.5}$ \\
a0.5h0.02rIR20k0.5 & 0.5 & 0.02 & 20.0 & 10$^{-0.5}$ \\
\hline
\hline
\end{tabular}
\caption{\label{tab:parameters} Physical Model Parameters.}
\end{table}

\section{Modelling the SED}
Vincent \& LeBohec \cite{vin10} have developed a Monte-Carlo simulation that allows, for a given set of parameters 
(magnetic field strength, background radiation...) a quantitative study of the energy distribution 
of the electrons in the gap and the associated electromagnetic radiation. 
In the following we demonstrate that the vacuum gap model is capable of reproducing the observed VHE 
spectrum at low and high activity states.

\begin{figure*}[h]
\begin{center}
\includegraphics[width=1\linewidth]{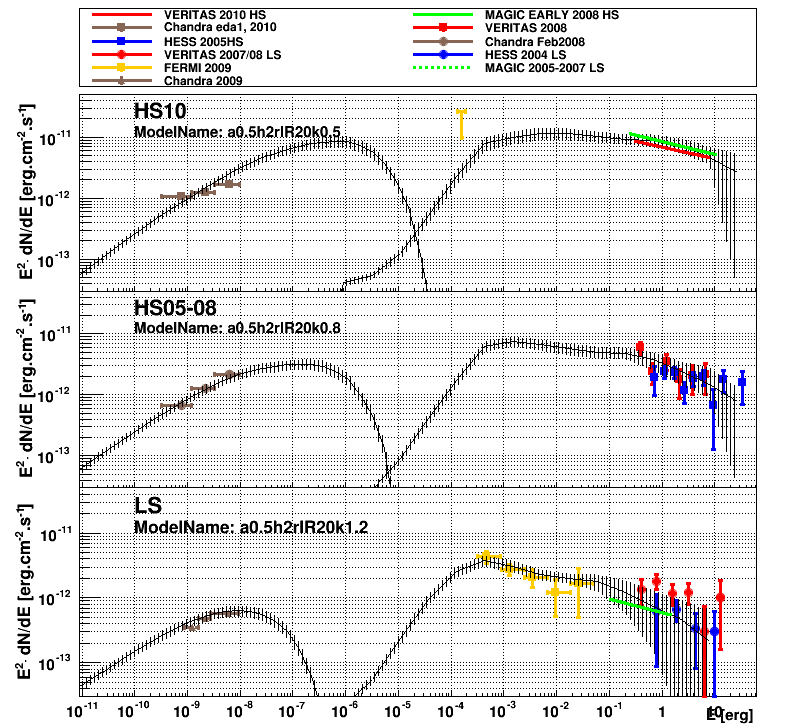}
\caption{Model outputs for the spectral energy distribution. 
The observed emission includes:
(1) Chandra observation in 2008, labelled with the solid filled brown squares with error bars,
Chandra observation in 2009 (solid filled brown triangles with error bars), and 
Chandra observation on 2010 April 1 (solid filled brown circles with error bars);
(2) Fermi-LAT observation in 2009, indicated with the filled squares with error bars~\cite{abd09}, and 
Fermi-LAT upper limit during the 2010 flare indicated by an orange arrow (http://fermisky.blogspot.de/2010/02/lat-limit-on-m87-during-tev-flare.html);
(3) HESS observations in 2004 and 2005, which are shown with the filled blue circles and the filled blue squares with error bars~\cite{aha06}; 
(4) MAGIC best-fit spectrum for data taken between 2005 and 2007 shown in green dash-dotted curve~\cite{ale12}, and on 2008 February 1 in green curve~\cite{alb08};
(5) VERITAS spectrum between 2007 December and 2008 May represented by filled red circles, and during the 2008 February flare 
represented by red filled squares~\cite{acc10}; the fit function reported during the 2010 April flare shown in red line~\cite{ali12}.
}
\label{overalSED}
\end{center}
\end{figure*}

%\subsection{2005, 2008 \& 2010 high states}
The radio galaxy M87 is normally a steady and weak source at TeV energies and 
to date, there have been three well documented flaring episodes: 2005 April, 
2008 February and 2010 April. 
The spectra measured during the TeV flares did not show any evidence of a cut-off, 
suggesting a weak $\gamma\gamma$ absorption. 
The long-term monitoring observations afford the opportunity of a more detailed look into the characteristics of 
the source at low emission level.
In figure \ref{overalSED}, we show both the spectral energy distribution (SED) of the synchrotron and the 
inverse Compton scattered radiation. 
We report (black lines) the emission from the magnetosphere model relative to the high states observed in 
2005-2008 (denoted HS05-08) and 2010 (denoted HS10), together with the low emission state (denoted LS). 
The data from different VHE instruments have also been reproduced: 
The fit functions reported by MAGIC in early 2008 February (green line; \cite{alb08}), 
and between 2005 and 2007 - characteristics to a low-emission state - 
(green dash-dotted lines; \cite{ale12}), the fit function reported by VERITAS 
during the 2010 April flare (red line; \cite{ali12}); 
the differential energy spectra measured by VERITAS \cite{acc09} during the 2008 February flare 
(red filled squares) and between 2007 December and 2008 May - corresponding to a low state emission - (red filled circles); 
the differential energy spectrum obtained from the 2004 (red filled circles) and the 2005 (blue filled squares) HESS data \citep{aha06}. 
In the high-energy gamma-ray range, the orange filled squares correspond to the Fermi-LAT energy spectrum from Abdo \textit{et al.}~\cite{abd09}, 
the orange arrow corresponds to the 95$\%$ confidence upper limits ($>100$~MeV) reported by Fermi-LAT and coincident with 
the announcement by MAGIC of a TeV flare in 2010 February
(http://fermisky.blogspot.de/2010/02/lat-limit-on-m87-during-tev-flare.html).
The brown filled squares reproduced the X-ray emission as measured by Chandra in 2008 and 
are contemporaneous to the joint MAGIC, HESS and VERITAS 2008 observation campaign, 
the brown filled circles report the nuclear emission on 2010 April 11 - this event was observed at 
a time when the TeV flaring was over~\cite{har11}, 
the brown filled triangles reproduce the images obtained in $\sim6$ week intervals between 
2008 and 2009 May observed (PI: D.~E.~Harris).

%%%The model is completely specified by the following four parameters: (i) the luminosity of the background radiation 
%%%field $L_{\mathrm{IR}}$, (ii) the energy of the seed photons for the IC scattering $E_{\mathrm{IR}}$ 
%%%(iii) the local value of the magnetic field and (iv) the rotation-induced electric field. 
%%%The adopted parameters are reported in Table \ref{parameters}.  

The magnetosphere model provides a satisfactory solution for both the high and the low states observed. 
It is compatible with the initial assumption that the different phases of activity are originating from the 
same emission region near the radio core. 
And the strength of the induced electric field  drives the acceleration efficiency of the charged particles. 
A feature of the model is that the TeV and X-ray fluxes are correlated. 
It should be emphasised that the 2010 April data are not simultaneous, in particular, 
by the night when Chandra obtained the first observation, the TeV flaring was over. 
%And 
This may explain the over-estimated X-ray flux level predicted by the magnetosphere model. \\

\subsection{Estimate of the local value of the charge density}
\label{section:Estimate of the local value of the charge density}
The longitudinal electric field in the accelerating gap arises due to deviation of the charge density from the local GJ charge density. 
When the absolute value of the local charge density $\rho$ is less than $\rho_{\mathrm{GJ}}$, an accelerating electric field is formed, leading to the cascade development. 
When |$\rho_{\mathrm{GJ}}$| < |$\rho$|, a decelerating electric field is formed suppressing the particle acceleration in the gap.
%%%At densities below the Goldreich-Julian density, the electric field can directly accelerated particles to high Lorentz factors. 
Therefore the magnetosphere model is consistent if the local value of the charged density is weak enough to be responsible for the 
development of the instabilities. 
In this section, we calculate the Goldreich-Julian density and the local charge density.  
We then attempt to evaluate the accretion rate.   

Estimate of the accretion rate $\dot{m}$ in an inefficient accretion flow is about a factor $\alpha$ lower than in a Bondi 
flow $\dot{m}\sim\alpha\dot{m}_{\mathrm{Bondi}}$, where $\alpha\sim0.1$ is the viscosity parameter \citep{nar02}, so that one expects  
the Goldreich-Julian density $n_{\mathrm{GJ}}$ to be $\sim 5\times10^{-6}~\mathrm{cm}^{-3}$. 
Using the Poisson's equation
\begin{equation}
\label{poisson eq.}
\nabla^2 V=4\pi e(n_{\pm}-n_{\mathrm{GJ}}),
%%%\nabla.\mathbf{E}=e\frac{n_{\pm}-n_{\mathrm{GJ}}}{\epsilon_0},
\end{equation}
one finds the local value of the density $n_{\pm}$ is related to the critical density $n_{\mathrm{GJ}}$ by
\begin{equation}
\label{number density}
|n_{\pm}-n_{\mathrm{GJ}}|\sim1.2\times 10^{24}\kappa B_{0}\frac{R_{\mathrm{G}}^2}{r^4}~\mathrm{cm}^{-3}
\end{equation}
More details are included in the Appendix~\ref{First appendix}.
%For our standard model with $\kappa=10^{-0.5}$, $B_{0}\sim15~\mathrm{G}, r\sim R_{\mathrm{G}}$, so $n_{\pm}\sim2\times10^{-6}~\mathrm{cm}^{-3}.$
For our standard model, 
we consider the strength of the magnetic field in the vicinity of the black hole and in the vacuum gap to be respectively $15~\mathrm{G}$ and $0.1~\mathrm{G}$. 
For a split-monopole field geometry, the magnetic field scales as $\propto r^{-2}$, which suggests the distance of the vacuum gap to the black hole to be  $\sim 12\times R_{G}$. 
Assuming $n_{\mathrm{GJ}}\sim r^{-2}$, the critical charge density $n_{\mathrm{GJ}}$ at the gap's location is $\sim 3.5\times10^{-8}~\mathrm{cm}^{-3}$, 
and for $\kappa=10^{-0.5}$, the local value of the charge density $n_{\pm}\sim3.4\times 10^{-8}~\mathrm{cm}^{-3}$.

%/n_{\mathrm{GJ}}$ to be $\sim0.01$. % to be $\sim4\times10^{-6}~\mathrm{cm}^{-3}$.  
%%%This estimate does not satisfy the condition $n_{\pm}=n_{\mathrm{GJ}}$, and a gap will form where the charged 
%%%particles accelerate.
%The condition $n_{\pm}=n_{\mathrm{GJ}}$ is not satisfied, and a gap will form where the charged 
%particles accelerate.
%The seed charge density $n_{\mathrm{\pm}}$ may also be related to the accretion rate $\dot{m}$ through equation (\ref{n_pm}), 
%and one finds that $\dot{m}\sim6\times 10^{-4}$.
%This value is smaller than the critical accretion rate derived in $\S$\ref{section:The role of the inefficient accretion flow}, 
%and confirms a gap may form during a period of low accretion.
%We estimate the critical charge density $n_{\mathrm{GJ}}$ in the gap .
However, these estimates are uncertain because of their dependence on the accretion rate $\dot{m}$, the electron-ion temperature 
ratio, and the standard BZ predictions \citep{pen13}. 
But exploring these parameter dependencies are beyond the scope of this paper.

%%\begin{deluxetable}{rrrrrrrr}
%%\tablecolumns{8}
%%\tablewidth{0pc}
%%%\tablecaption{Input parameters of the models for the magnetosphere model shown in figures \ref{sedHS} \& \ref{sedLS}.}
%%\tablehead{
%%\colhead{}    &  \multicolumn{3}{c}{Non-shell Stars} &   \colhead{}   &
%%\multicolumn{3}{c}{Shell Stars} \\
%%\cline{2-4} \cline{6-8} \\
%%\colhead{} & \colhead{$n_{\mathrm{IR}}$[cm$^{-3}$]} & \colhead{B[G]} & \colhead{$\kappa$ (HS)} & \colhead{$\kappa$ (LS)}
%%\startdata
%%0.06 & $10^{12}$ & 0.03 & $10^{-2}$ & $10^{-3}$
%%\enddata
%%\end{deluxetable}
\section{TeV luminosity and the hole's spin}

Using the observational data in the VHE regime (i.e. $>100~$GeV), the $\gamma$-ray luminosity can be estimated. 
%%%Assuming a steady and isotropic emission, 
The photon spectral luminosity is given by \citep{geh97}
\begin{equation}
\label{luminosity}
%%%L_{\gamma}(E_{\gamma})=4\pi D_{L}^{2}\frac{\Phi_{\gamma}(E_{\gamma})}{(1+z)^2}e^{\tau(E_{\gamma}, z)},
%%%l_{\gamma}(E_{\gamma})=\eta D_{L}^{2}\frac{\Phi_{\gamma}(E_{\gamma})}{(1+z)^2}e^{\tau(E_{\gamma}, z)},
l_{\gamma}(E_{\gamma})=2\pi(1-\cos\theta) D_{L}^{2}\frac{\Phi_{\gamma}(E_{\gamma})}{(1+z)^2}e^{\tau(E_{\gamma}, z)},
\end{equation}
here $\theta$ is the opening angle of the emission cone, $\Phi_{\gamma}(E_{\gamma})$ is the differential photon spectrum in units of photons per unit time, unit area and energy $E_{\gamma}$, 
$\tau$ represents the $\gamma\gamma$ optical depth for a photon traveling from the source with initial energy $E_{\gamma}$. 
The values used for $\tau$ were taken from \cite{fra08}. 
The luminosity distance to the source $D_{L}$ is defined as
\begin{equation}
D_{L}(z)=(1+z)\int_0^z cdz/H(z)\sim 5.6\times 10^{25}~\mathrm{cm},
\end{equation}
where $H(z)$ is the Hubble parameter at redshift $z$. 
Corrections for both adiabatic energy losses due to the redshift and attenuation due to interactions with the 
radiation are considered in equation \ref{luminosity}. 

For the differential photon spectrum detected at Earth, we use the values of the fit parameter measured by \cite{ali12}
\begin{equation}
\Phi_{\gamma}(E_{\gamma})_{E_{\gamma}>350~\mathrm{GeV}}=\Phi_0\left(\frac{E_{\gamma}}{\mathrm{TeV}}\right)^{\Gamma},
\end{equation}
where the flux normalisation constant $\Phi_0$ is $4.71\times10^{-12}~\mathrm{cm^{-2}s^{-1}TeV^{-1}}$ and the spectral index $\Gamma$ is 2.19. 
The measured integral luminosity $L_{\gamma}^{\mathrm{meas.}}(E_{\gamma}>350~\mathrm{GeV})$ is obtained by integrating $l_{\gamma}(E_{\gamma})$, and 
\begin{equation} 
L_{\gamma}^{\mathrm{meas.}}(E_{\gamma}>350~\mathrm{GeV})\simeq 4\times10^{41}(1-\cos\theta)~\mathrm{erg.s}^{-1}.
%%%7\times10^{40}\eta~\mathrm{erg.s}^{-1}. 
\end{equation}

%%%The rate of energy output arising from the BZ process by a black hole of angular momentum $a$, with a magnetic field $B$ 
%%%normal to the horizon of the hole, is given by \citep{mac82, gho97}
%%%}
%%%\begin{equation}
%%%L_{\mathrm{BZ}}=\frac{1}{32}\frac{\Omega_{\mathrm{F}}(\Omega_{\mathrm{H}}-\Omega_{\mathrm{F}})}{\Omega_{\mathrm{H}}^2}
%%%a^2B^2R_{\mathrm{H}}^2c
%%%\end{equation}
%%%\textbf{
%%%where $R_{\mathrm{H}}$ is the horizon radius and $\Omega_{\mathrm{H}}$ (resp. $\Omega_{\mathrm{F}}$) is the angular velocity of 
%%%the hole (resp. the field lines). 
%%%The power output is maximized for $\Omega_{\mathrm{F}}=1/2\Omega_{\mathrm{H}}$ and this is what we adopt here. 
%%%Using $R_{\mathrm{H}}\simeq2GM_{\mathrm{BH}}/c^2$ and for the magnetic field strength close to $R_{\mathrm{H}}\sim 15~\mathrm{G}$, 
%%%introduced in Section \ref{section:The role of the inefficient accretion flow}, yields
%%%}
%%%\begin{equation}
%%%L_{\mathrm{BZ}}\simeq 5.3\times 10^{40}a^2~\mathrm{erg.s}^{-1}.
%%%\end{equation}

For each electron propagating in the gap of height $h$, the energy-loss rate is estimated as 
$dE/dt\simeq ecV/h$. 
The electric potential difference across a gap, with a magnetic field $B$, generated by a black hole of angular momentum $a$,  
can be expressed as,
%%%\begin{equation}
\begin{eqnarray}
V\simeq1.47\times 10^{18}a(1+\sqrt{1-a^2})\left(\frac{B}{10^4~\mathrm{G}}\right)
\left(\frac{M}{10^9~\mathrm{M}_{\odot}}\right)
\left(\frac{h}{R_{\mathrm{G}}}\right)^2~\mathrm{statvolt}
%%%\end{equation}
\end{eqnarray}
Inside the gap, the field aligned electric field is unscreened and this implies that the density of electrons 
in the gap is limited by $n_{\mathrm{GJ}}$. 
The funnel is a strongly magnetised region that develops over the poles, and is nearly devoid of matter. %evacuated, 
%The geometry of the region 
Its geometry consists of a cone with an opening $\theta$.

%Taking into account that the volume of the gap is roughly $h^3$, 
Taking into account that the volume of the gap is $(\pi/3)R_{\mathrm{G}}^3\times (12^3-10^3)\tan^{2}\theta$, 
the maximum gamma-ray power that can be produced by charged particles accelerating in the gap, 
regardless of the specific gamma-ray production mechanism is
\begin{eqnarray}
%&&L_{\gamma}\simeq n_{\pm}h^3\tan^{2}\theta\cdot(dE/dt)\nonumber\\
&&L_{\gamma}\simeq 7.6\times 10 ^{47} n_{\pm}\tan^{2}\theta\cdot(dE/dt)\nonumber\\
&&\phantom{L_{\gamma}}\simeq 8.1\times 10^{51} n_{\pm}\tan^2 \theta\cdot a(1+\sqrt{1-a^2})\left(\frac{B}{10^4~\mathrm{G}}\right)
%%&&\phantom{L_{\gamma}}\simeq 2\times 10^{19} n_{\pm}h^{2}\tan^2 \theta\cdot a(1+\sqrt{1-a^2})\left(\frac{B}{10^4~\mathrm{G}}\right)
\left(\frac{M}{10^9~\mathrm{M}_{\odot}}\right)\left(\frac{h}{R_{\mathrm{G}}}\right)^2\nonumber\\
%&\phantom{L_{\gamma}}\simeq 2\times 10^{19} n_{\pm} h^{2}\tan^2\theta\cdot a(1+\sqrt{1-a^2})\left(\frac{B}{10^4~\mathrm{G}}\right)
%&&\phantom{L_{\gamma}}\left(\frac{M}{10^9~\mathrm{M}_{\odot}}\right)\left(\frac{h}{R_{\mathrm{G}}}\right)^2\nonumber\\
%&&\phantom{L_{\gamma}}\simeq 4\times 10^{26} \dot{m}^4 h^{2}\tan^2\theta\cdot a(1+\sqrt{1-a^2})\left(\frac{B}{10^4~\mathrm{G}}\right)
%\left(\frac{M}{10^9~\mathrm{M}_{\odot}}\right)\left(\frac{h}{R_{\mathrm{G}}}\right)^2\nonumber\\
&&\phantom{L_{\gamma}}\simeq 6.6\times 10^{40}\tan^{2}\theta\cdot a(1+\sqrt{1-a^2})~\mathrm{erg.s^{-1}},
\end{eqnarray}
where $V$ is the voltage drop across the gap height $h=2R_{\mathrm{G}}$, $n_{\pm}\sim3.4\times10^{-8}~\mathrm{cm}^{-3}$ and $B\sim10^{-1}-10^{-2}~\mathrm{G}$.

\begin{figure}[tbp]
\centering % \begin{center}/\end{center} takes some additional vertical space
\includegraphics[width=0.7\textwidth]{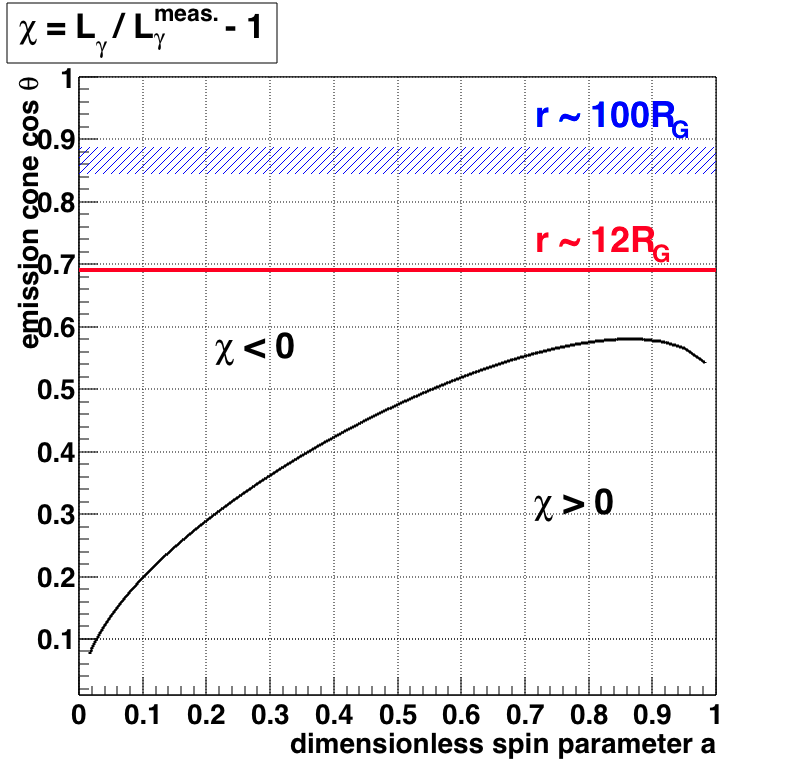} %allowedParameters.png}
\caption{
\label{fig:contour} Determination of the allowed parameters $(\theta,~a)$ satisfying the condition $L_{\gamma}/L_{\gamma}^{\mathrm{meas.}}-1>0$. 
The confidence interval represents the jet opening angle on scales 0.04pc which corresponds to $\sim100R_{\mathrm{G}}$~\cite{jun99}. 
The red horizontal line indicates the jet opening angle, $\tan\theta\sim 0.5\theta_{\mathrm{w}}/D$, at an apparent core distance of $D\sim12R_{\mathrm{G}}$ derived from~\cite{doe12}.
}
\end{figure}

Figure \ref{fig:contour} shows the allowed region that satisfies the condition $L_{\gamma}>L_{\gamma}^{\mathrm{meas.}}$.
Thus, if the angular momentum of the BH is large enough, and the TeV $\gamma$-ray emission beamed into a cone with an opening 
angle $\theta\gtrsim 55^{\circ}$, the non-thermal power of the gap can account for the observed TeV $\gamma$-ray luminosity. 
VLBI observations~\cite{jun99} show that the jet opening angle is continuously increasing as the core is approached, reaching $\sim60^{\circ}$ within 0.04pc 
($\sim 100R_{\mathrm{G}}$). 
Furthermore, Doeleman \textit{et al.}~\cite{doe12} resolves the base of the M87 jet. 
They show that the projected jet width fits with a power law $\theta_{\mathrm{w}}\propto r^{0.69}$, where $\theta_{\mathrm{w}}$ is the jet width and $r$ is the separation from the core. 
At an apparent core distance of $1 R_{\mathrm{G}}$, the result of the fit gives $\theta_{\mathrm{w}}\sim4.5 R_{\mathrm{G}}$. 
For comparison purposes, we find the jet width value to be $\sim25 R_{\mathrm{G}}$ at a distance of $\sim12 R_{\mathrm{G}}$, i.e. $\theta \sim \tan^{-1}(0.5 \cdot \theta_{\mathrm{w}}/12)\sim 46^{\circ}$. 
Our estimate is consistent with those based on radio observations for spin parameters $a\gtrsim0.6$.

% which seems to indicate a flaw in the magnetosphere model especially at low spin parameter $a$. 
%The result of the fit Assuming the jet width to be $~4.5 R_{\mathrm{G}}$ at an apparent core distance of $~1 R_{\mathrm{G}}$, we find the jet width value to be $\sim25 R_{\mathrm{G}}$
%at an apparent core distance of $\sim12 R_{\mathrm{G}}$, i.e. $\theta\sim\atan(0.5\cdot\theta_{\mathrm{w}}/12)\sim46^{\circ}$. 
%For comparison, the jet width value is $\sim25 R_{\mathrm{G}}$.
%
%However, our estimate is based on a possible vacuum gap located closer to the core, and the possibility remains that the two luminosities $L_{\gamma}$ and $L_{\gamma}^{\mathrm{meas.}}$ may be reconciled if 
%the opening angle of the jet is getting wider near the core. 
%A better estimate for $n_{\mathrm{GJ}}$ using general relativistic magnetohydrodynamics simulation~\cite{mos11} gives 
%$n_{GJ}\sim0.3\dot{m}^{1/2}m_{8}^{-3/2}$~cm$^{-3}\sim10~cm$^{-3}$, and $\theta\gtrsim 85^{\circ}$,
 
%
%\label{Estimate of the local value of the charge density}} is uncertain and 

%This estimate is in relatively good agreement with ones %those 
%based on radio observations of the inner regions~\cite{jun99}.   

\section{Summary}

Nearby under-luminous and mis-aligned AGN emerge as prime candidate sources where non-thermal VHE processes occurring close to the central BH are observable. 
The radio galaxy M87 is a unique source that harbours a substantially inclined jet 
$\sim(15-25)^{\circ}$ with respect to the observer. 
The broadband emission originating near the supermassive black hole is not necessarily 
swamped by the relativistically boosted jet flux, and the high-energy emission signatures, e.g. the day-scale variability, have triggered new theoretical developments. 

The particular picture developed in this paper considers a %funnel 
region which is magnetically dominated but not, in 
general, in a state of force-free equilibrium. 
The fluctuations - initiated by the intermittence of the accretion rate - in the outflow will be 
responsible for the formation of gaps. 
The charges injected into the gap are quickly accelerated by the large potential drop, 
and reach large Lorentz factor at which energy gain is balanced by synchrotron and inverse 
Compton losses.
The exact location of the vacuum gap in the BH magnetosphere can only be found by numerically modelling the charge supply, the geometry of the accretion flow, and the formation of the force-free magnetosphere. 
Therefore, the gap model examined in this study should be considered a useful toy model for the gap geometry. 

In the case of M87, we show that the 2008 and 2010 observed TeV $\gamma$-ray emission can be 
explained by electron acceleration and $\gamma$-ray production in a compact region close to the event horizon. 
The same mechanism, with a relatively weak rotation-induced electric fields, may also account for the 2009 
low state. 
In this scenario, the intermittences of the gap potential are related to the changes of the charge density into the 
magnetosphere. 
We estimated these densities for the flare and the low states and we noted that the changes 
will be associated with the accretion rate fluctuations.

The magnetosphere model is applicable to variable TeV-emitting AGN that accrete at sufficiently low rates, in particularly  Cen A. 
If Cen A hosts a radiatively inefficient inner accretion disk, it could thus be another primary in which VHE processes close to the 
black hole may allow a fundamental diagnosis of its immediate environment, offering important insights into the central engine 
in AGNs.
%%\textbf{FIND A NICE CONCLUSION - READ RIEGER\&AHARONIAN}
%%\begin{figure}[!h]
%%\epsscale{.90}
%%\plotone{plots/sedLowState.eps}
%%\caption{.}
%%\label{sedLowState}
%%\end{figure} 

\acknowledgments
%%%I thank the anonymous referee for helpful suggestions which improved the manuscript.
I wish to thank Stephan LeBohec for hepful discussions. 
I am grateful to Michelle Hui for her comments on the VERITAS observations 
and Dan Harris who provided the Chandra data. 
I acknowledge support through the Helmholtz Alliance for Astroparticle Particle.

%\paragraph{Note added.} This is also a good position for notes added
%after the paper has been written.

\appendix
\section{Derivation of equation \ref{number density}}
\label{First appendix}
The electric potential difference across a gap height $h$ generated by a maximally rotating BH is~\cite{tho86}
\begin{eqnarray}
%V\simeq4.41\times 10^{20}\kappa \left(\frac{B}{10^4~\mathrm{G}}\right)
V\simeq1.47\times 10^{18}\kappa \left(\frac{B}{10^4~\mathrm{G}}\right)
\left(\frac{M}{10^9~\mathrm{M}_{\odot}}\right)
\left(\frac{h}{R_{\mathrm{G}}}\right)^2~\mathrm{statvolt}
\end{eqnarray}
For a split-monopole field geometry, the magnetic field scales as $\propto r^{-2}$. 
Assuming that the gap height $h$ extends along the magnetic field lines to the height $h=2R_{\mathrm{G}}$, 
the potential drop and the divergence of the electric field parallel to the magnetic field are given by
\begin{eqnarray}
\nabla^2 V&&\sim 7 \times 10^{15}\kappa B_{0}\frac{R_{\mathrm{G}}^{2}}{r^{4}}~\mathrm{statvolt.cm}^{-2},
%%%V&&\sim 10^{18}\kappa B_{0}\left(\frac{R_{\mathrm{G}}}{r}\right)^{2}~\mathrm{volt},\\
%%%\nabla E&&\sim 2 \times 10^{18}\kappa B_{0}\frac{R_{\mathrm{G}}^{2}}{r^{4}}~\mathrm{volt.m}^{-2},
\label{eq1}
\end{eqnarray}
where $B_{0}$ is the magnetic field strength near of the horizon of the hole. 
We approximate the volume integral over the Poisson's equation~\ref{poisson eq.} by
\begin{eqnarray}
\label{eq2}
\int dx^3 \nabla^2 V &\approx& \nabla^2 V\cdot \int dx^3,\\
\int dx^3 4\pi e(n_{\pm}-n_{\mathrm{GJ}})&\approx& 4\pi e(n_{\pm}-n_{\mathrm{GJ}})\cdot \int dx^3.
\label{eq3}
\end{eqnarray}  
Using equations \ref{eq1}, \ref{eq2} \& \ref{eq3}, we get %then estimate the charge density,
 \begin{equation}
|n_{\pm}-n_{\mathrm{GJ}}|\sim 1.2\times10^{24}\kappa B_{0}\frac{R_\mathrm{G}^2}{r^4}~\mathrm{cm}^{-3}.
\end{equation}
%We consider the strength of the magnetic field in the vicinity of the black hole and in the vacuum gap to be respectively $15~\mathrm{G}$ and $0.1~\mathrm{G}$. 
%The distance $r$ of the gap to the black hole is $\sim 12\times R_{G}$. 
%We set $\kappa=1$, $n_{\pm}=0$, $B_{0}\sim 15~\mathrm{G}$, $r\sim 12\times R_{G}$, 
%and find $n_{\mathrm{GJ}}\sim 9 \times 10^{-10}~\mathrm{cm}^{-3}$. 
%Using the current density $J^{\mu}$ derived from the BZ monopole solution as given in~\cite{mck04}, 
%Mo{\'s}cibrodzka \textit{et al.}~\cite{mos11} notice that $n_{\mathrm{GJ}}\sim 1/r^3$, and 
%the critical charge density in the vicinity of the black hole is $~2 \times 10^{-6}~\mathrm{cm}^{-3}$. 
%This is consistent with the estimate given in \S\ref{section:Estimate of the local value of the charge density} .
% The bibliography will probably be heavily edited during typesetting.
% We'll parse it and, using the arxiv number or the journal data, will
% query inspire, trying to verify the data (this will probalby spot
% eventual typos) and retrive the document DOI and eventual errata.
% We however suggest to always provide author, title and journal data:
% in short all the informations that clearly identify a document.


\begin{thebibliography}{99}
%%%%%%%%%%
% \bibitem{a}
% Author, \emph{Title}, \emph{J. Abbrev.} {\bf vol} (year) pg.

\bibitem{geb09}Gebhardt, K., \& Thomas, J., \emph{The Black Hole Mass, Stellar Mass-to-Light Ratio, and Dark Halo in M87}, \emph{\apj} {\bf 700} (2009) [arXiv:0906.1492]
\bibitem{mei07}Mei, S., et al., \emph{The ACS Virgo Cluster Survey. XIII. SBF Distance Catalog and the Three-dimensional Structure of the Virgo Cluster}, \emph{\apj} {\bf 655} (2007) [astro-ph/0702510 ]
\bibitem{aha03}Aharonian, F., et al., \emph{Is the giant radio galaxy M87 a TeV gamma-ray emitter?}, \emph{\aaj} {\bf 403} (2003) [astro-ph/0302155]
\bibitem{aha06}Aharonian, F., et al., \emph{Fast Variability of Tera-Electron Volt $\gamma$ Rays from the Radio Galaxy M87}, \emph{Science} {\bf 314} (2006) [astro-ph/0612016]
\bibitem{har06} Harris, D.~E., Cheung, C.~C., Biretta, J.~A., Sparks, W.~B., Junor, W., Perlman, E.~S., \& Wilson, A.~S., \emph{The Outburst of HST-1 in the M87 Jet}, \emph{\apj} {\bf 640} (2006) [astro-ph/0511755]
\bibitem{har09} Harris, D.~E., Cheung, C.~C., Stawarz, {\L}., Biretta, J.~A., \& Perlman, E.~S., \emph{Variability Timescales in the M87 Jet: Signatures of E$^{2}$ Losses, Discovery of a Quasi Period in HST-1, and the Site of TeV Flaring}, \emph{\apj} {\bf699} (2009) [arXiv:0904.3925]
\bibitem{che07} Cheung, C.~C., Harris, D.~E., \& Stawarz, {\L}., \emph{Superluminal Radio Features in the M87 Jet and the Site of Flaring TeV Gamma-Ray Emission}, \emph{\apjl} {\bf663} (2007) [arXiv:0705.2448]
\bibitem{har11} Harris, D.~E., Massaro, F., Cheung, C.~C., et al., \emph{An Experiment to Locate the Site of TeV Flaring in M87}, \emph{\apj} {\bf743} (2011) [arXiv:1111.5343]
\bibitem{bei08} Beilicke, M., Hui, C.~M., Mazin, D., Raue, M., Wagner, R.~M., \& Wagner, S., \emph{A joint H.E.S.S./MAGIC/VERITAS observation campaign on the radio galaxy M 87 to probe the origin of VHE {$\gamma$}-ray emission}, \emph{American Institute of Physics Conference Series} {\bf1085} (2008)
\bibitem{acc09} Acciari, V.~A., Aliu, E., Arlen, T., et al., \emph{Radio Imaging of the Very-High-Energy {$\gamma$}-Ray Emission Region in the Central Engine of a Radio Galaxy}, \emph{Science} {\bf325} (2009) [arXiv:0908.0511]
\bibitem{bla77} Blandford, R.~D., \& Znajek, R.~L., \emph{Electromagnetic extraction of energy from Kerr black holes}, \emph{\mnras} {\bf179} (1977)
\bibitem{wil01} Wilms, J., Reynolds, C.~S., Begelman, M.~C., et al., \emph{XMM-EPIC observation of MCG-6-30-15: direct evidence for the extraction of energy from a spinning black hole?}, \emph{\mnras} {\bf328} (2001) [astro-ph/0110520]
\bibitem{mar03} Maraschi, L., \& Tavecchio, F., \emph{The Jet-Disk Connection and Blazar Unification}, \emph{\apj} {\bf593} (2003) [astro-ph/0205252]
\bibitem{geo05} Georganopoulos, M., Perlman, E.~S., \& Kazanas, D., \emph{Is the Core of M87 the Source of Its TeV Emission? Implications for Unified Schemes}, \emph{\apjl} {\bf634} (2005) [astro-ph/0510783]
\bibitem{geo03} Georganopoulos, M., \& Kazanas, D., \emph{Relativistic and Slowing Down: The Flow in the Hot Spots of Powerful Radio Galaxies and Quasars}, \emph{\apjl} {\bf589} (2003) [astro-ph/0304256]
\bibitem{geo05} Georganopoulos, M., Perlman, E.~S., \& Kazanas, D., \emph{Is the Core of M87 the Source of Its TeV Emission? Implications for Unified Schemes}, \emph{\apjl} {\bf634} (2005) [astro-ph/0510783]
\bibitem{tav08} Tavecchio, F., \& Ghisellini, G., \emph{Spine-sheath layer radiative interplay in subparsec-scale jets and the TeV emission from M87}, \emph{\mnras} {\bf385} (2008) [arXiv:0801.0593]
\bibitem{len08} Lenain, J.-P., Boisson, C., Sol, H., \& Katarzy{\'n}ski, K., \emph{SSC scenario for VHE emission from 2 radiogalaxies: M 87 and Cen A}, \emph{\aaj} {\bf478} (2008) [arXiv:0807.2733]
\bibitem{gia10} Giannios, D., Uzdensky, D.~A., \& Begelman, M.~C., \emph{Fast TeV variability from misaligned minijets in the jet of M87}, \emph{\mnras} {\bf402} (2010) [arXiv:0907.5005]
\bibitem{bar10} Barkov, M.~V., Aharonian, F.~A., \& Bosch-Ramon, V., \emph{Gamma-ray Flares from Red Giant/Jet Interactions in Active Galactic Nuclei}, \emph{\apj} {\bf724} (2010) [arXiv:1005.5252]
\bibitem{jun99} Junor, W., Biretta, J.~A., \& Livio, M., \emph{Formation of the radio jet in M87 at 100 Schwarzschild radii from the central black hole}, \emph{\nat} {\bf401} (1999)
\bibitem{ly07} Ly, C., Walker, R.~C., \& Junor, W., \emph{High-Frequency VLBI Imaging of the Jet Base of M87} \emph{\apj} {\bf660} (2007) [astro-ph/0701511]
\bibitem{asa12} Asada, K., \& Nakamura, M., \emph{The Structure of the M87 Jet: A Transition from Parabolic to Conical Streamlines}, \emph{\apjl} {\bf745} (2012) [arXiv:1110.1793]
\bibitem{ner07} Neronov, A., \& Aharonian, F.~A., \emph{Production of TeV Gamma Radiation in the Vicinity of the Supermassive Black Hole in the Giant Radio Galaxy M87}, \emph{\apj} {\bf671}  (2007) [arXiv:0704.3282]
\bibitem{rie08} Rieger, F.~M., \& Aharonian, F.~A., \emph{Production of TeV Gamma Radiation in the Vicinity of the Supermassive Black Hole in the Giant Radio Galaxy M87} \emph{International Journal of Modern Physics D} {\bf17} (2008) [arXiv:0704.3282]
\bibitem{vin10} Vincent, S., \& LeBohec, S., \emph{Monte Carlo simulation of electromagnetic cascades in black hole magnetosphere}, \emph{\mnras} {\bf409} (2010)
\bibitem{lev11} Levinson, A., \& Rieger, F., \emph{Variable TeV Emission as a Manifestation of Jet Formation in M87?}, \emph{\apj} {\bf730} (2011) [arXiv:1011.5319]
\bibitem{mac82} MacDonald, D., \& Thorne, K.~S., \emph{Black-hole electrodynamics - an absolute-space/universal-time formulation}, \emph{\mnras} {\bf198} (1982)
\bibitem{bes92} Beskin, V.~S., Istomin, Y.~N., \& Parev, V.~I., \emph{Filling the Magnetosphere of a Supermassive Black-Hole with Plasma}, \emph{\sovast} {\bf36} (1992)
\bibitem{hir98} Hirotani, K., \& Okamoto, I., \emph{Pair Plasma Production in a Force-free Magnetosphere around a Supermassive Black Hole}, \emph{\apj} {\bf497} (1998)
\bibitem{lev00} Levinson, A., \emph{Particle Acceleration and Curvature TeV Emission by Rotating, Supermassive Black Holes}, \emph{Physical Review Letters} {\bf85} (2000)
\bibitem{aha02} Aharonian, F.~A., Belyanin, A.~A., Derishev, E.~V., Kocharovsky, V.~V., \& Kocharovsky, V.~V., \emph{Constraints on the extremely high-energy cosmic ray accelerators from classical electrodynamics}, \emph{\prd} {\bf66} (2002) [astro-ph/0202229]
\bibitem{wal74} Wald, R.~M., \emph{Black hole in a uniform magnetic field}, \emph{\prd} {\bf10} (1994)
\bibitem{gol69} Goldreich, P., \& Julian, W.~H., \emph{Pulsar Electrodynamics}, \emph{\apj} {\bf157} (1969)
\bibitem{haw06} Hawley, J.~F., \& Krolik, J.~H., \emph{Magnetically Driven Jets in the Kerr Metric}, \emph{\apj} {\bf641} (2006) [astro-ph/0512227]
\bibitem{bec09} Beckwith, K., Hawley, J.~F., \& Krolik, J.~H., \emph{Transport of Large-Scale Poloidal Flux in Black Hole Accretion}, \emph{\apj} {\bf707} (2009) [arXiv:0906.2784]
\bibitem{kha04} Kharb, P., \& Shastri, P., \emph{Optical nuclei of radio-loud AGN and the Fanaroff-Riley divide}, \emph{\aaj} {\bf425} (2004) [astro-ph/0401042]
\bibitem{why04} Whysong, D., \& Antonucci, R., \emph{Thermal Emission as a Test for Hidden Nuclei in Nearby Radio Galaxies}, \emph{\apj} {\bf602} 116 (2004) [astro-ph/0207385]
\bibitem{mac97} Macchetto, F., Marconi, A., Axon, D.~J., et al., \emph{The Supermassive Black Hole of M87 and the Kinematics of Its Associated Gaseous Disk}, \emph{\apj} {\bf489} (1997) [astro-ph/9706252]
\bibitem{dim03} Di Matteo, T., Allen, S.~W., Fabian, A.~C., Wilson, A.~S., \& Young, A.~J., \emph{Accretion onto the Supermassive Black Hole in M87}, \emph{\apj} {\bf582} (2003) [astro-ph/0202238]
\bibitem{nar95} Narayan, R., \& Yi, I., \emph{Advection-dominated Accretion: Underfed Black Holes and Neutron Stars}, \emph{\apj} {\bf452} (1995) [astro-ph/9411059]
\bibitem{mos11} Mo{\'s}cibrodzka, M., Gammie, C.~F., Dolence, J.~C., \& Shiokawa, H., \emph{Pair Production in Low-luminosity Galactic Nuclei}, \emph{\apj} {\bf735} (2011) [arXiv:1104.2042]
\bibitem{mck04} McKinney, J.~C., \& Gammie, C.~F., \emph{A Measurement of the Electromagnetic Luminosity of a Kerr Black Hole}, \emph{\apj} {\bf611} (2004) [astro-ph/0404512]
\bibitem{kom04} Komissarov, S.~S., \emph{Electrodynamics of black hole magnetospheres}, \emph{\mnras} {\bf350} (2004)
\bibitem{kom02} Komissarov, S.~S., \emph{Time-dependent, force-free, degenerate electrodynamics}, \emph{\mnras} {\bf336} (2002) [astro-ph/0202447]
\bibitem{mck06} McKinney, J.~C., \emph{General relativistic magnetohydrodynamic simulations of the jet formation and large-scale propagation from black hole accretion systems
}, \emph{\mnras} {\bf368} (2006) [astro-ph/0603045]
\bibitem{con13} Contopoulos, I., Kazanas, D., \& Papadopoulos, D.~B., \emph{The Force-free Magnetosphere of a Rotating Black Hole}, \emph{\apj} {\bf765} (2013) [arXiv:1212.0320]
\bibitem{alb08} Albert, J., Aliu, E., Anderhub, H., et al., \emph{Very High Energy Gamma-Ray Observations of Strong Flaring Activity in M87 in 2008 February}, \emph{\apjl} {\bf685} (2008) [arXiv:0806.0988]
\bibitem{ale12} Aleksi{\'c}, J., Alvarez, E.~A., Antonelli, L.~A., et al., \emph{MAGIC observations of the giant radio galaxy M 87 in a low-emission state between 2005 and 2007}, \emph{\aaj} {\bf544} (2012) [arXiv:1207.2147]
\bibitem{ali12} Aliu, E., Arlen, T., Aune, T., et al., \emph{VERITAS Observations of Day-scale Flaring of M 87 in 2010 April}, \apj {\bf746} (2012)
\bibitem{acc09} Acciari, V.~A., Aliu, E., Arlen, T., et al., \emph{Veritas 2008-2009 Monitoring of the Variable Gamma-ray Source M 87}, \emph{Science} {\bf325} (2009) [arXiv:1005.0367]
\bibitem{abd09} Abdo, A.~A., Ackermann, M., Ajello, M., et al., \emph{Fermi Large Area Telescope Gamma-Ray Detection of the Radio Galaxy M87}, \emph{\apj} {\bf707} (2009) [arXiv:0910.3565]
\bibitem{har11} Harris, D.~E., Massaro, F., Cheung, C.~C., et al., \emph{An Experiment to Locate the Site of TeV Flaring in M87}, \emph{\apj} {\bf743} (2011) [arXiv:1111.5343]
\bibitem{nar02} Narayan, R., \emph{Lighthouses of the Universe: The Most Luminous Celestial Objects and Their Use for Cosmology}, {\bf405} (2002)
\bibitem{pen13} Penna, R.~F., Narayan, R., \& Sadowski, A., emph{General relativistic magnetohydrodynamic simulations of Blandford-Znajek jets and the membrane paradigm}, \emph{\mnras} {\bf436} (2013) [arXiv:1307.4752]
\bibitem{alb08} Albert, J., Aliu, E., Anderhub, H., et al., \emph{Very High Energy Gamma-Ray Observations of Strong Flaring Activity in M87 in 2008 February}, \emph{\apjl}, {\bf685} (2008)
\bibitem{acc10} Acciari, V.~A., Aliu, E., Arlen, T., et al., \emph{Veritas 2008-2009 Monitoring of the Variable Gamma-ray Source M 87}, \emph{\apj}, {\bf716} (2010)
\bibitem{geh97} Gehrels, N., \emph{Use of {$\nu$}F$_{\nu}$ spectral energy distributions for multiwavelength astronomy.}, \emph{Nuovo Cimento B Serie} {\bf112} (1997)
\bibitem{fra08} Franceschini, A., Rodighiero, G., \& Vaccari, M., \emph{Extragalactic optical-infrared background radiation, its time evolution and the cosmic photon-photon opacity}, \emph{\aaj}, {\bf487} (2008)
\bibitem{tho86} Thorne, K.~S., Price, R.~H., \& MacDonald, D.~A.\ 1986, Black Holes: The Membrane Paradigm,
\bibitem{doe12} Doeleman, S.~S., Fish, V.~L., Schenck, D.~E., et al.\ 2012, Science, 338, 355
% Please avoid comments such as "For a review'', "For some examples",
% "and references therein" or move them in the text. In general,
% please leave only references in the bibliography and move all
% accessory text in footnotes.

% Also, please have only one work for each \bibitem.


\end{thebibliography}
\end{document}